\documentclass[a4paper,11pt]{article}
\pdfoutput=1 

\usepackage{jcappub} 
 
\usepackage[utf8]{inputenc}
\usepackage[english]{babel}
\usepackage[T1]{fontenc}
\usepackage{bbm}
\usepackage[margin=2.5cm]{geometry}               
\usepackage{amssymb}
\usepackage{amsmath}	
\newcommand\numberthis{\addtocounter{equation}{1}\tag{\theequation}}

\usepackage{graphicx}	
\usepackage{wrapfig}
\usepackage{braket}
\usepackage{stackengine,scalerel}

\title{\boldmath Mimetic K-essence}

\author[a]{Pavel Jirou\v{s}ek,}
\author[b]{Keigo Shimada,}
\author[c]{Alexander Vikman,}
\author[b]{and Masahide Yamaguchi}

\affiliation[a]{\it High Energy Physics, Cosmology \& Astrophysics Theory Group, University of Cape Town,\\Private Bag, Cape Town 7700, South Africa}
\affiliation[c]{\it Department of Physics, Tokyo Institute of Technology,\\Tokyo 152-8551, Japan}
\affiliation[b]{\it CEICO - Central European Institute for Cosmology and Fundamental Physics,\\ FZU - Institute of Physics of the Czech Academy of Sciences,\\Na Slovance 1999/2, 18220 Prague 8, Czech Republic}

\emailAdd{pavel.jirousek@uct.ac.za}
\emailAdd{shimada.k.ah@m.titech.ac.jp}
\emailAdd{vikman@fzu.cz}
\emailAdd{gucci@phys.titech.ac.jp}

\abstract{We propose a new non-trivial way to combine mimetic dark matter with the mimetic formulation of unimodular gravity. This yields a Weyl-invariant higher-derivative scalar-vector-tensor theory. We demonstrate that on-shell its behavior mimics GR with an additional k-essence scalar. The overall scale of the k-essence arises as an integration constant - a global degree of freedom. Interestingly, we find that the resulting fluid cannot make transition through ultra-relativistic equation of state. We develop a method to find a mimetic theory corresponding to any eligible k-essence and identify, which k-essences can or cannot be reproduced this way. Finally, we show that abandoning the Weyl symmetry of the setup allows us to obtain both unimodular gravity and mimetic dark matter simultaneously, from one conformal redefinition of the metric.}

\begin{document}
\maketitle
\flushbottom
\section{Introduction}
One of the prime goals of modifying gravity\footnote{For recent reviews, see e.g. \cite{Clifton:2011jh,Heisenberg:2018vsk}.} is to provide viable models of \emph{dark matter} (DM) and/or \emph{dark energy} (DE) as geometric extensions of general relativity (GR). In view of the so-called coincidence problem\footnote{Indeed, the energy densities of DM and DE are of the same order now, despite their completely different behaviour with the redshift, see e.g. \cite{Sahni:2004ai}.} it would be elegant to combine both dark entities under one roof. 

A theory of particular interest to us is the \emph{mimetic dark matter} \cite{Chamseddine:2013kea}, in which GR is modified by proposing a composite structure to the \emph{physical metric} $g_{\mu\nu}$ which determines the geodesics in the spacetime. Thus, in the \emph{mimetic} DM, $g_{\mu\nu}$ is no longer considered as an independent variable. The form of the composite structure of $g_{\mu\nu}$ proposed in \cite{Chamseddine:2013kea} is given by
\begin{equation}
g_{\mu\nu}=h_{\mu\nu}\,\,h^{\rho\sigma}\partial_{\rho}\phi\,\partial_{\sigma}\phi\ ,\label{mimetic DM ansatz}
\end{equation}
where a \emph{auxiliary metric} $h_{\mu\nu}$ is introduced along with the \emph{mimetic scalar} $\phi$ as new independent variables. Applying this ansatz to the Einstein-Hilbert action with usual matter fields, as a seed theory, produces an additional matter sector equivalent to a fluid-like irrotational dust \cite{Chamseddine:2013kea,Golovnev:2013jxa,Barvinsky:2013mea}. Hence, the mimetic scalar field $\phi$ plays the role of the velocity potential.  
This is a simple realization of cold dark matter valid on cosmologically large scales where linear perturbations are applicable and where one can ignore caustics. 

An important feature of the ansatz \eqref{mimetic DM ansatz} is the invariance of the physical metric $g_{\mu\nu}$ (and of the resulting theory) with respect to the Weyl transformations of the dynamical variable $h_{\mu\nu}$:
\begin{equation}
    h_{\mu\nu}\rightarrow \omega^{2}(x)\,h_{\mu\nu}\ .\label{sec1: weyl transformation}
\end{equation}
This Weyl-invariant scalar-tensor setup has generated a significant amount of followup research, for a non-exhaustive list of references see e.g. \cite{Chamseddine:2014vna,Deruelle:2014zza,Mirzagholi:2014ifa,Capela:2014xta,Ramazanov:2015pha,Arroja:2015wpa, Arroja:2015yvd, Arroja:2017msd, Hammer:2015pcx,Chamseddine:2016ktu,Chamseddine:2016uef, Chamseddine:2016uyr, Babichev:2017lrx,Takahashi:2017pje, Hirano:2017zox, Ganz:2018mqi,Ganz:2018vzg,Chamseddine:2018gqh, Langlois:2018jdg, Chamseddine:2018qym,Ganz:2019vre, Chamseddine:2019gjh, Ramazanov:2016xhp,Zlosnik:2018qvg,Jirousek:2022rym,Golovnev:2022jts,Jirousek:2022jhh} and \cite{Sebastiani:2016ras} for a review. Quite surprisingly, in view of the results of \cite{Deruelle:2014zza} we have demonstrated in recent work \cite{Jirousek:2022rym} that Weyl-invariance of the composite physical metric is not necessary to realize the mimetic DM. However, the Weyl invariance guaranties the absences of other branches of solutions, which can be dust-like as well as those with spacelike gradients of the scalar field. Yet, each branch of the solutions corresponds to a Weyl-invariant theory.  
This mimetic construction inspired other generalizations of ansatz \eqref{mimetic DM ansatz} where the physical metric $g_{\mu\nu}$ is given as a conformal rescaling of a dynamical variable metric $h_{\mu\nu}$
\begin{equation}
    g_{\mu\nu}=\Omega\, h_{\mu\nu}\ ,\label{sec1: general mimetic}
\end{equation}
where $\Omega$ transforms as $\Omega\rightarrow \omega^{-2}\,C$ preserving the Weyl-invariance of the physical metric. One of the first such generalizations \cite{Gorji:2018okn,Gorji:2019ttx} has considered  
\begin{equation}
    \Omega=\sqrt{F_{\mu\nu}F^{\mu\nu}}\ ,\label{sec1: gorji ansatz}
\end{equation}
where $F_{\mu\nu}$ is the gauge field strength for a Yang-Mills gauge field. This results in the introduction of a matter component mimicking the spatial curvature.
While, another Yang-Mills invariant involving the Hodge dual of the field strength tensor, 
\begin{equation}
\widetilde{F}^{\mu\nu}=\frac{1}{2}\,\frac{\varepsilon^{\mu\nu\alpha\beta}}{\sqrt{-h}}\,F_{\alpha\beta}\,,\label{Hodge}
\end{equation}
can be utilized \cite{Hammer:2020dqp} 
\begin{equation}
\Omega=\sqrt{F_{\mu\nu}\widetilde{F}^{\mu\nu}} \,,\label{axionic_Omega}
\end{equation}
to yield \emph{mimetic} DE as a new formulation of the covariant version of the unimodular gravity \cite{Henneaux:1989zc}. The relation with \cite{Henneaux:1989zc} is even closer, if one considers \cite{Jirousek:2018ago} conformal factor 
\begin{equation}
\Omega=\sqrt{\nabla_{\mu}V^{\mu}}\,,\quad\text{where}\quad\nabla_{\mu}V^{\mu}=\frac{1}{\sqrt{-h}}\partial_{\mu}\left(\sqrt{-h}V^{\mu}\right)\,,
\label{Omega_Vector}
\end{equation}
instead of \eqref{axionic_Omega}. In this proposal the Weyl transformation needs to be extended to the vector as $V^{\mu}\rightarrow\omega^{-4}\,V^{\mu}$ in order to preserve the symmetry.
We have demonstrated \cite{Hammer:2020dqp} that both substitutions \eqref{axionic_Omega} and \eqref{Omega_Vector} are equivalent and are just more suitable to establish connection either to the high energy physics \eqref{axionic_Omega} or to the Henneaux-Teitelboim covariant unimodular gravity \eqref{Omega_Vector}. We will use these substitutions interchangeably. 

It is straightforward to see that any linear combination of the conformal factors $h^{\rho\sigma}\partial_{\rho}\phi\,\partial_{\sigma}\phi$ from \eqref{mimetic DM ansatz} and \eqref{sec1: gorji ansatz} and \eqref{axionic_Omega} (or \eqref{Omega_Vector}) yields a new conformal factor with the correct transformation properties, i.e. with conformal weight 2. We can use such construction to generate new Weyl invariant theories applying the ansatz \eqref{sec1: general mimetic} to any seed theory. In fact, the space of the possible expressions for mimetic conformal factor is even larger, e.g. the coefficients of the linear combination can be functions of the ratios of the individual conformal factors and still retain the desired Weyl invariance. Following this observation we ask ourselves a natural question: can we unite both components of the dark sector in a single mimetic theory by considering such combinations of \eqref{mimetic DM ansatz} and \eqref{axionic_Omega}?
In this work we pursue this question. We explore the dynamics that result from the most general Weyl invariant ansatz \eqref{sec1: general mimetic} using any admissible combination of \eqref{mimetic DM ansatz} and \eqref{axionic_Omega}. We lay down the details of the construction in section \ref{section: mixing}. In section \ref{section: EoM} we derive the equations of motion and demonstrate the appearance of an additional integration constant. In section \ref{section: mimetic k-essence} we show that this theory is (with very rare exceptions) on-shell equivalent to k-essence \cite{Armendariz-Picon:1999hyi,Chiba:1999ka,Armendariz-Picon:2000nqq,Armendariz-Picon:2000ulo} with an extra global dynamical degree of freedom. It is well-known \cite{Greiter:1989qb,Son:2000ht,Son:2002zn} that, for timelike gradients of the scalar field, k-essence describes (super)fluid hydrodynamics\footnote{For most recent discussions including breaking of shift-symmetry, see \cite{Kourkoulou:2022doz,Nicolis:2022llw}.}. Thus, combining fluid-like dust with arbitrary cosmological constant we obtained a superfluid with an arbitrary energy scale. The global degree of freedom represents the overall energy scale of the k-essence Lagrangian and arises as a constant of integration. Similarly to our previous work \cite{Jirousek:2020vhy}, we argue that this global degree of freedom is canonically conjugated to a quantity measuring normalised action between Cauchy hypersurfaces. 

We provide a general method that allows us to find a mimetic theory that reproduces a given k-essence in section  \ref{section: reconstruction}. In the process, we discover that not all k-essences can be fully recovered through this scheme. Indeed, the mimetic description breaks down once the equation of state of the k-essence becomes that of radiation. We discuss this in section \ref{section: breakdown}. Finally, we discuss the role of the underlying Weyl symmetry in section \ref{section: non-Weyl}. We show that abandoning this requirement yields an additional non-dynamical degree of freedom in our theory, which can be associated with the conformal mode of the auxiliary metric $h_{\mu\nu}$. In a similar fashion to \cite{Jirousek:2022jhh}, this conformal mode can be integrated out, which yields two branches of solutions. One where the Weyl symmetry is restored and a second corresponding to the unmodified seed theory. In contrast to \cite{Jirousek:2022jhh}, we find a special case of our construction where the Weyl symmetry is not restored. These are equivalent to mimetic DM with a potential \cite{Chamseddine:2014vna} which was first introduced in before the mimetic construction in \cite{Lim:2010yk}. Interestingly, in this branch both mimetic DM and unimodular gravity (mimetic DE) can be realized simultaneously, through a single transformation of the metric which we present.

\section{Mimetic mixing}\label{section: mixing}
In this work we propose a mimetic ansatz that involves two terms that transform homogeneously under the Weyl group: First, the kinetic term of a scalar field:
\begin{equation}
Y=h^{\mu\nu}\partial_{\mu}\phi\partial_{\nu}\phi.\label{kinetic term}
\end{equation}
and second, the Chern-Pontryagin invariant of a gauge field $A_{\mu}$
\begin{equation}
D=F_{\mu\nu}\widetilde{F}^{\mu\nu}=\frac{1}{2\sqrt{-h}}\epsilon^{\mu\nu\rho\sigma}D_{\mu}A_{\nu}\cdot D_{\rho}A_{\sigma}.\label{Pontryagin invariant}
\end{equation}
Here $\epsilon^{\mu\nu\sigma\rho}$ is the Levi-Civita symbol and the dot represents taking the trace. The details of the associated gauge group are not too important in this setting. The kinetic term has a conformal weight 2 and on its own produces the mimetic dark matter \cite{Chamseddine:2013kea}, while the Chern-Pontryagin invariant has weight 4 and results in unimodular gravity \cite{Hammer:2020dqp}. We consider the most general ansatz involving the above terms
\begin{equation}
g_{\mu\nu}=h_{\mu\nu}\Omega(Y,D,\phi).\label{mimetic ansatz}
\end{equation}
The invariance with respect to the gauge group corresponding to $A_{\mu}$ restricts any direct dependence on the gauge fields themselves, however, an explicit dependence on $\phi$ is admissible. The overall Weyl invariance of the right hand side is ensured by imposing that the function $\Omega$ has a conformal weight 2. In other words, it satisfies the following algebraic property
\begin{equation}
\Omega(\omega^{-2}Y,\omega^{-4}D,\phi)=\omega^{-2}\Omega(Y,D,\phi).\label{Omega transformation}
\end{equation}
This can be alternatively characterized through a generalization of the Euler's homogeneous function theorem as
\begin{equation}
Y \,\Omega_{Y}+2D \,\Omega_{D}=\Omega.\label{Euler homogeneous theorem}
\end{equation}
The terms \eqref{kinetic term} and \eqref{Pontryagin invariant} will often appear evaluated with the physical metric instead of the original $h_{\mu\nu}$. To distinguish these instances we introduce additional notation
\begin{align}
    X&=g^{\mu\nu}\partial_{\mu}\phi\partial_{\nu}\phi,\\
    P&=F_{\mu\nu}F^{\star}\,\!^{\mu\nu}=\frac{1}{2\sqrt{-g}}\epsilon^{\mu\nu\rho\sigma}D_{\mu}A_{\nu}\cdot D_{\rho}A_{\sigma}.
\end{align}
This work relies heavily upon manipulation of the arguments of the function $\Omega$. In order to capture these, without cluttering the equations, we adopted a shorthand notation, where the partial derivatives also tell us about the arguments of the differentiated functions. That is
\begin{align*}
    \Omega_{Y}&\equiv\frac{\partial\Omega}{\partial Y}(Y,D,\phi),\qquad\qquad\qquad
    \Omega_{X}\equiv\frac{\partial\Omega}{\partial X}(X,P,\phi),\\
    \Omega_{D}&\equiv\frac{\partial\Omega}{\partial D}(Y,D,\phi),\qquad\qquad\qquad
    \Omega_{P}\equiv\frac{\partial\Omega}{\partial P}(X,P,\phi),
\end{align*}
and similarly for other functions. Furthermore, the instances of $\Omega$ with no argument specified are meant to be evaluated using the original variables $Y$ and $D$: 
\begin{equation}
    \Omega\equiv\Omega(Y,D,\phi).
\end{equation}
Unfortunately, the above notation does not cover all our needs and we will at times resort to specifying the arguments directly, which then takes precedence.

Note that unlike $Y$ and $D$, the composite variables $X$ and $P$ are not independent of each other. Indeed, the second immediate implication of \eqref{Omega transformation} is the appearance of a \textit{mimetic constraint} \cite{Golovnev:2013jxa}, which introduces a relation between the two. We find this relation by simply evaluating the mimetic factor using $X$ and $P$:
\begin{equation}
\Omega(X,P,\phi)=\Omega(\Omega^{-1}Y,\Omega^{-2}D,\phi)=\Omega^{-1}\Omega=1.
\end{equation}
In total we get
\begin{equation}
\Omega(X,P,\phi)=1.\label{mimetic constraint}
\end{equation}
The appearance of such constraint is a recurring theme in mimetic theories and is a first sign of the constrained nature of the theory. This constraint comes back as a Lagrange multiplier when we turn to gauge invariant variables \cite{Hammer:2015pcx}.\

In this work we obtain the action for our theory by plugging the mimetic ansatz \eqref{mimetic ansatz} into the Einstein-Hilbert action $S_{EH}[g]$ with an unspecified matter Lagrangian $S_{matter}[g,\Psi_{M}]$. We only assume that the matter sector does not explicitly depend on $A_{\mu}$ and $\phi$. Thus our action is
\begin{equation}
S_{mim}[h,A,\phi,\Psi_{M}]=S_{EH}[g(h,A,\phi)]+S_{matter}[g(h,A,\phi),\Psi_{M}].\label{mimetic theory}
\end{equation}
The above theory is a higher order scalar-vector-tensor theory due to the higher derivative nature of the Einstein-Hilbert term. Indeed, expanding the scalar curvature in the gravitational part of the action highlights this structure\footnote{We are using the signature convention $(+,-,-,-)$ and the units $c=\hbar=1$, $M_{pl}=(8\pi G)^{-1/2}=1$}:
\begin{equation}
    S_{EH}[g(h,A,\phi)]=-\frac{1}{2}\int d^{4}x\sqrt{-h}\bigg [\Omega R(h)+\frac{3}{2}\frac{h^{\mu\nu}\partial_{\mu}\Omega\partial_{\nu}\Omega}{\Omega}\Bigg].\label{action expanded}
\end{equation}
Written in this form the Lagrangian contains derivatives of the mimetic factor $\Omega$, which itself contains derivatives of $\phi$ and $A_{\mu}$, resulting in higher derivatives. One can reduce the order of the derivatives by introducing an extra scalar field $\theta$ constrained via a Lagrange multiplier $\lambda$ as
\begin{equation}
    S_{EH}[g(h,A_{\mu},\phi),\theta,\lambda]=-\frac{1}{2}\int d^{4}x\sqrt{-h}\bigg [\theta R(h)+\frac{3}{2}\frac{h^{\mu\nu}\partial_{\mu}\theta\partial_{\nu}\theta}{\theta} +\lambda\big (\Omega-\theta\big )\Bigg].\label{action expanded constraint}
\end{equation}
The Weyl invariance of the theory can be preserved by equipping the fields $\theta$ and $\lambda$ with conformal weights $2$:
\begin{equation}
    \theta\rightarrow \omega^{-2}\theta,\qquad\lambda\rightarrow\omega^{-2}\lambda,\qquad\mathrm{as}\qquad h_{\mu\nu}\rightarrow\omega^{2}h_{\mu\nu}.
\end{equation}
This allows us to introduce the following Weyl invariant variables
\begin{align}
    g_{\mu\nu}&=\theta h_{\mu\nu},\\\label{field redefinition g}
    \tilde{\lambda}&=\lambda\theta^{-1},\\
    A_{\mu}&=A_{\mu},\\
    \phi&=\phi.\label{field redefinition phi}
\end{align}
Performing these redefinitions in \eqref{action expanded constraint} $\theta$ disappears entirely. The Lagrangian becomes that of standard GR with mimetic constraint \eqref{mimetic constraint} enforced through a Lagrange multiplier $\tilde{\lambda}$
\begin{equation}
    S[g,\theta,\phi,A_{\mu},\lambda]=-\frac{1}{2}\int d^{4}x\sqrt{-g}\bigg [R(g) +\tilde{\lambda}\big (\Omega(X,P,\phi)-1\big )\Bigg]\,.\label{action mimetic constraint}
\end{equation}
One can further alternate the action by introducing another scalar field $\pi$ and a Lagrange multiplier $q$

\begin{equation}
    S[g,\theta,\phi,A_{\mu},\lambda,q,\pi]=-\frac{1}{2}\int d^{4}x\sqrt{-g}\left(R(g)+\lambda\left[\Omega(X,\pi,\phi)-1\right]-q\left[F_{\mu\nu}F^{\star}\,\!^{\mu\nu}-\pi\right]\right)\,,\label{action two constraints}
\end{equation}
where the Hodge dual is defined through \eqref{Hodge} with the determinant of the physical metric $g_{\mu\nu}$. This form of the action is quadratic in derivatives of the gauge field and is a convenient starting point to investigate canonical structure of the theory, as well as for potential axionic UV-completions, similar to our previous works \cite{Hammer:2020dqp,Jirousek:2020vhy}.

\section{Equations of motion}\label{section: EoM}
Let us derive the equations of motion associated to the mimetic fields. These fields enter exclusively through the physical metric and therefore the first step of the variation is always
\begin{equation}
\delta S_{mim}=-\frac{1}{2}\int d^{4}x\sqrt{-g}(G_{\mu\nu}-T_{\mu\nu})\delta g^{\mu\nu},\label{einstein variation}
\end{equation}
where both the Einstein tensor $G_{\mu\nu}$ and the energy momentum tensor $T_{\mu\nu}$ are defined with reference to the $g_{\mu\nu}$ only. The inverse physical metric can be obtained by inverting relation \eqref{mimetic ansatz}
\begin{equation}
g^{\mu\nu}=\frac{h^{\mu\nu}}{\Omega(Y,D,\phi)}.
\end{equation}
Note that $h^{\mu\nu}$ and $g^{\mu\nu}$ are inverses to their covariant counterparts $h_{\mu\nu}$ and $g_{\mu\nu}$. Varying this relation yields
\begin{equation}
\delta_{h}g^{\mu\nu}=\Omega^{-1}\Big [\delta^{\mu}_{\rho}\delta^{\nu}_{\sigma}-h^{\mu\nu}\frac{\Omega_{Y}}{\Omega}\partial_{\rho}\phi\partial_{\sigma}\phi-\frac{1}{2}h^{\mu\nu}\frac{\Omega_{D}}{\Omega}D h_{\rho\sigma}\Big ]\delta h^{\rho\sigma}.
\end{equation}
We notice that every term in the square bracket is Weyl invariant\footnote{It can be easily checked that the derivatives $\Omega_{Y}$ and $\Omega_{D}$ have conformal weights $0$ and $-2$ respectively.} and thus we can express it through the physical metric as
\begin{equation}
\delta_{h}g^{\mu\nu}=\Omega^{-1}\Big [\delta^{\mu}_{\rho}\delta^{\nu}_{\sigma}-g^{\mu\nu}\Omega_{X}\partial_{\rho}\phi\partial_{\sigma}\phi-\frac{1}{2}g^{\mu\nu}\Omega_{P}P g_{\rho\sigma}\Big ]\delta h^{\rho\sigma}.
\end{equation}
The instance of $\Omega$ in the denominators vanish due to the mimetic constraint \eqref{mimetic constraint}. Plugging this into \eqref{einstein variation} and dividing the $\Omega^{-1}$ out, we obtain the modified Einstein equation
\begin{equation}
G_{\mu\nu}-(G-T)\Omega_{X}\partial_{\mu}\phi\partial_{\nu}\phi-\frac{G-T}{2}\Omega_{P}Pg_{\mu\nu}=T_{\mu\nu}.\label{einstein equation}
\end{equation}
This equation is now fully Weyl invariant as all quantities here are evaluated on the physical metric.\ 

Before we continue let us briefly comment on the above equation. Our theory \eqref{mimetic theory} is inherently Weyl invariant and as such it cannot produce unique solutions for $h_{\mu\nu}$. The physical information in $h_{\mu\nu}$ is carried by the physical metric $g_{\mu\nu}$ and therefore one can at this step forgo $h_{\mu\nu}$ and think about  equation \eqref{einstein equation} as an equation for $g_{\mu\nu}$ directly.

The original Weyl invariance has one further consequence - one equation is missing as the trace part of \eqref{einstein equation} vanishes identically. In other words, the information about the conformal mode is absent. This is the case even for the physical metric $g_{\mu\nu}$. We can see this directly by taking the trace
\begin{equation}
(G-T)\big (1-\Omega_{X}X-\Omega_{P}2P\big )=0,
\end{equation}
which is satisfied trivially due to \eqref{Euler homogeneous theorem} and \eqref{mimetic constraint}. The missing information is instead encoded into the mimetic constraint \eqref{mimetic constraint}, which has to be added to the equations of motion in order to form a complete set.

We now focus on the equations of motion of the other mimetic fields $A_{\mu}$ and $\phi$. These fields are themselves Weyl invariant and thus their equations of motion are invariant as well. The variation of \eqref{mimetic ansatz} with respect to $\phi$ and $A_{\mu}$ yields:
\begin{equation}
\delta g_{\mu\nu}=h_{\mu\nu}\Big [2\Omega_{Y}h^{\rho\sigma}\partial_{\rho}\phi\partial_{\sigma}\delta\phi+\Omega_{\phi}(Y,D,\phi)\delta\phi+\Omega_{D}\widetilde{F}^{\rho\sigma}D_{\rho}\delta A_{\sigma}\Big ].
\end{equation}
The expression in the brackets has conformal weight 2, which exactly balances the weight of the auxiliary metric $h_{\mu\nu}$. Thus we switch to the physical metric everywhere and write
\begin{equation}
\delta g_{\mu\nu}=g_{\mu\nu}\Big [2\Omega_{X}g^{\rho\sigma}\partial_{\rho}\phi\partial_{\sigma}\delta\phi+\Omega_{\phi}(X,P,\phi)\delta\phi+\Omega_{P}F^{\star\rho\sigma}D_{\rho}\delta A_{\sigma}\Big ]\, ,
\end{equation}
where $F^{\star\mu\nu}$ is again defined through \eqref{Hodge} with the determinant of the physical metric $g_{\mu\nu}$. Plugging this into \eqref{einstein variation} we obtain the equations of motion for $\phi$
\begin{equation}
\nabla_{\mu}\Big [(G-T)\Omega_{X}\partial^{\mu}\phi\Big ]=\frac{G-T}{2}\Omega_{\phi},\label{eom phi}
\end{equation}
and for $A_{\mu}$
\begin{equation}
F^{\star}\,\!^{\mu\nu}\partial_{\nu}\big [(G-T)\Omega_{P}\big ]=0.\label{eom A}
\end{equation}
At this point both equations are evaluated with the physical metric only. The last equation implies
\begin{equation}
(G-T)\Omega_{P}=-2Q=const \,,\label{eom A Q}
\end{equation}
as long as $F^{\star}\,\!^{\mu\nu}$ is invertible as a matrix \cite{Hammer:2020dqp}. It is easy to check that this is the case if and only if $P\neq 0$, which we will assume from now on.\

At this point we can easily recover various cosmological fluids in the current setup. We can see that the equation \eqref{einstein equation} is an Einstein equation with an additional perfect fluid component with a velocity potential $\phi$, and energy density and pressure
\begin{align}
    \rho_{mim}&=(G-T)\Big (\Omega_{X}X+\Omega_{P}P/2\Big ),\\
    p_{mim}&=-(G-T)\Omega_{P}P/2 \,.
\end{align}
Assuming a constant $w$ parameter $p_{mim}=w\rho_{mim}$ yields
\begin{equation}
    (2\Omega_{X}X+\Omega_{P}P)w=-\Omega_{P}P.
\end{equation}
Applying the Euler homogeneous theorem \eqref{Euler homogeneous theorem} allows us to eliminate $\Omega_{X}X$ to obtain
\begin{equation}
    \Omega=\frac{3w-1}{2w}\Omega_{P}P.
\end{equation}
The solution of this equation is straightforward and yields
\begin{equation}
    \Omega(X,P,\phi)=P^{\frac{2w}{3w-1}}F(X,\phi).
\end{equation}
Furthermore the $X$ dependence can be fixed from the \eqref{Omega transformation} to obtain the final form
\begin{equation}
    \Omega(X,P,\phi)=X\Big (\frac{P}{X^{2}}\Big )^{\frac{2w}{3w-1}}F(\phi).\label{mimetic w fluid}
\end{equation}
The above construction clearly fails when $w=1/3$ or $w=0$. The first one is a first indication of a deeper incompatibility of the mimetic scenario with the ultra-relativistic equation of state, which will be addressed in detail later in this paper. The case $w=0$ can be easily circumvented by eliminating $\Omega_{P}P$ in favor of $\Omega_{X}X$ instead. In the end it yields the same result \eqref{mimetic w fluid}.

\section{Mimetic K-essence}\label{section: mimetic k-essence}
The mimetic constraint \eqref{mimetic constraint} can be viewed as an implicit equation that can (in principle) be solved for $P$.\footnote{The conditions for the existence of such solution are given from the implicit function theorem and require that $\Omega_{D}|_{\Omega=1}\neq 0$.} Such a solution may be written formally as
\begin{equation}
P=P(X,\phi),\label{implicit solution}
\end{equation}
where $P(X,\phi)$ is a general function of two variables satisfying
\begin{equation}
    \Omega(X,P(X,\phi),\phi)=1,
\end{equation}
By plugging this solution\footnote{Strictly speaking it is necessary to find a direct solution for the gauge field $A_{\mu}$ whose Pontryagin class reproduces the sought after solution for $P$. However, such solution can always be found as long as the gauge group contains SU(2) as a subgroup and the spacetime is globally hyperbolic} into the rest of the equations of motion we obtain scalar-tensor equations that are of second order in derivatives and have only up to first order derivatives of the scalar field in the equation \eqref{einstein equation} - a situation very reminiscent of k-essence! And indeed, it turns out that every solution that allows for \eqref{implicit solution} is equivalent to GR with k-essence. The overall scale of the corresponding k-essence action in our setting becomes a global degree of freedom.\

Let us make this correspondence concrete. We start by considering a solution \eqref{eom A Q} and use it to solve for $G-T$.
\begin{equation}
G-T=-\frac{2Q}{\Omega_{P}}.
\end{equation}
We can immediately see that vanishing $Q$ implies $G=T$, thus our theory contains the unmodified seed theory as a special case. Plugging this into the modified Einstein equation \eqref{einstein equation} we obtain
\begin{equation}
G_{\mu\nu}+2Q\frac{\Omega_{X}}{\Omega_{P}}\partial_{\mu}\phi\partial_{\nu}\phi+QPg_{\mu\nu}=T_{\mu\nu},\label{einstein equation K}
\end{equation}
and analogously from equation \eqref{eom phi} we get
\begin{equation}
2Q\nabla_{\mu}\Big [\frac{\Omega_{X}}{\Omega_{P}}g^{\mu\nu}\partial_{\nu}\phi\Big ]=Q\frac{\Omega_{\phi}}{\Omega_{P}}.\label{eom phi K}
\end{equation}
At this point we assume that we have some solution \eqref{implicit solution}. Differentiating the mimetic constraint evaluated on such solution with respect to $X$ yields
\begin{equation}
\Omega_{X}+\Omega_{P}P_{X}=0,
\end{equation}
which can be used to express
\begin{equation}
P_{X}=-\frac{\Omega_{X}}{\Omega_{P}}.
\end{equation}
Similarly we get
\begin{equation}
P_{\phi}=-\frac{\Omega_{\phi}}{\Omega_{P}}.
\end{equation}
Plugging these into \eqref{einstein equation K} and \eqref{eom phi K} we obtain
\begin{align}
G_{\mu\nu}-2QP_{X}\partial_{\mu}\phi\partial_{\nu}\phi+QPg_{\mu\nu}&=T_{\mu\nu},\\
2Q\nabla_{\mu}\Big [P_{X}g^{\mu\nu}\partial_{\nu}\phi\Big ]&=QP_{\phi}.
\end{align}
Which are clearly equations of motion of GR with k-essence described by the Lagrangian
\begin{equation}
    \mathcal{L}(X,\phi)=QP(X,\phi)\,.\label{mimetic k essence}
\end{equation}
The overall scale $Q$ has been obtained as an integration constant in this setting rather then as a given parameter and represents the above mentioned global degree of freedom. 
It is instructive to inspect the action \eqref{action two constraints} from where it follows that $Q=q/2$.
It is important that the sign of $Q$ can be both positive and negative. However, this sign never changes during evolution. This sign just defines the physically relevant i.e. stable regions of the phase space.
For instance, the sign of $Q$ differentiates between usual ghost condensate \cite{Arkani-Hamed:2003pdi} and the inverted one \cite{Scherrer:2004au} which appear just as different branches in our mimetic theory. This appearance of the constant of integration in front of the k-essence Lagrangian resembles the situation in the so-called generalized unimodular gravity \cite{Barvinsky:2017pmm,Barvinsky:2019agh,Barvinsky:2020sxl}. 

Following the Faddeev-Jackiw procedure \cite{Faddeev:1988qp}, and comparing with our previous works \cite{Hammer:2020dqp,Jirousek:2020vhy} one can immediately infer\footnote{It is easier to see in the formulation with divergence of a vector \eqref{Omega_Vector}.} that $Q$ is canonically conjugated to a global quantity $s$ which measures spacetime average of $\pi$ between Cauchy hypersurfaces. 
\begin{equation}
s\left(t_{2}\right)-s\left(t_{1}\right)=\int d^{4}x\sqrt{-g}\,\pi\,.\label{conjugated}
\end{equation}
Crucially, at least on-shell, constraint enforced by $\lambda$ in \eqref{action two constraints} implies that $\pi=P(X,\phi)$, so that $s\left(t_{2}\right)-s\left(t_{1}\right)$ is proportional to the on-shell action of the corresponding k-essence theory. 

Interestingly, appearance of $Q$ renders the coupling constant of the gauge theory corresponding to $A_{\mu}$ irrelevant. Indeed, by field redefinition of $A_{\mu}$ one can always factor out the coupling constant from the field strength
\begin{equation}
    F_{\mu\nu}\rightarrow \frac{1}{g}F_{\mu\nu},
\end{equation}
and by doing so shifting $P$ as
\begin{equation}
    P\rightarrow\frac{1}{g^{2}}P.
\end{equation}
However, in the equation \eqref{mimetic k essence} this just corresponds to a redefinition of $Q$
\begin{equation}
    \mathcal{L}(X,\phi)=\frac{Q}{g^{2}}P(X,\phi)=\Tilde{Q}P(X,\phi).
\end{equation}
Thus all values of $g$ are equally capable of producing the same dynamics. This reasoning has further consequence on the space of mimetic theories of this type. 

In the above discussion we have considered a single solution \eqref{implicit solution} of the mimetic constraint \eqref{mimetic constraint}, however, since this solution arises from an implicit equation, there is in general no guarantee that there exists only a single such solution. Thus for the same pair of values $X,\phi$ several possible values of $P$ might be admissible. This means that the theory may support several k-essences governed by different Lagrangians. If we turn to quantum theory then such solutions may exist simultaneously in superposition!

\section{Reconstructing the $\Omega$ factor}\label{section: reconstruction}
So far we have shown that nearly all choices of $\Omega$ result in some form of k-essence, with the only exception being the mimetic dark matter scenario. However, it is not yet clear if all k-essences can be represented in this way. In this section we propose a simple method that allows us to construct mimetic theories engineered to reproduce a given k-essence. From the limitations of this procedure we are then able to identify which k-essences cannot be obtained from mimetic gravity.

Suppose we are given a k-essence Lagrangian $\mathcal{L}(X,\phi)$ that we wish to reproduce via mimetic gravity. In the previous section we have seen that such Lagrangian arises in our setup as a solution of the mimetic constraint for $P$. Thus, we are looking for a function $\Omega(X,P,\phi)$ with conformal weight 1 that yields
\begin{equation}
P=\mathcal{L}(X,\phi),\label{P of X}
\end{equation}
as a solution of the implicit equation
\begin{equation}
    \Omega(X,P,\phi)=1.
\end{equation}
The method for obtaining a corresponding $\Omega$ can be found rather easily. We assume that $X$ and $P$ arose from $Y$ and $D$ as
\begin{align*}
    P=\Omega^{-2}D,\\
    X=\Omega^{-1}Y,\numberthis{}\label{mimetic picture variables}
\end{align*}
as they do in a mimetic theory. For the moment we have no idea what $\Omega$ should be and thus, we treat it as an independent variable. Plugging these relations into \eqref{P of X} provides us with an implicit equation
\begin{equation}
\Omega^{-2}D=\mathcal{L}(\Omega^{-1}Y,\phi).\label{Omega implicit}
\end{equation}
The formal solution for $\Omega$ in terms of $Y,D$ and $\phi$  yields an expression of the correct form:
\begin{equation}
\Omega=\Omega(Y,D,\phi).\label{Omega solution}
\end{equation}
Crucially, this solution has two key properties that make it the sought after mimetic factor: first $\Omega=1$ yields \eqref{P of X}. This is a straight forward consequence as plugging $\Omega=1$ into \eqref{mimetic picture variables} and \eqref{Omega implicit} immediately reproduces \eqref{P of X}. Second, it satisfies the Euler homogeneous theorem \eqref{Euler homogeneous theorem}. This can be checked without actually finding the solution \eqref{Omega solution} as the derivatives of $\Omega$ can be found by differentiating the implicit equation \eqref{Omega implicit}. This yields
\begin{align*}
\Omega_{D}=\frac{1}{2\Omega \mathcal{L}(X,\phi)-\mathcal{L}_{X}(X,\phi)Y},\\
\Omega_{Y}=\frac{-\mathcal{L}_{X}(X,\phi)\Omega}{2\Omega \mathcal{L}(X,\phi)-\mathcal{L}_{X}(X,\phi)Y}.\numberthis{}
\end{align*}
Plugging these into \eqref{Euler homogeneous theorem} and using \eqref{Omega implicit} we get
\begin{equation}
2D \,\partial_{D}\Omega +Y \partial_{Y}\Omega=\Omega.
\end{equation}
Thus we see that $\Omega(Y,D,\phi)$ indeed satisfies the homogeneity conditions \eqref{Omega transformation}.\

Let us illustrate the above procedure in practise with an example. We consider a simple cuscuton \cite{Afshordi:2006ad} Lagrangian
\begin{equation}
    P=\mathcal{L}(X,\phi)=\sqrt{X}F(\phi).
\end{equation}
The first step is to recover the $\Omega$ dependence by using \eqref{mimetic picture variables} to obtain
\begin{equation}
    \Omega^{-2}D=\sqrt{\Omega^{-1}Y}F(\phi).
\end{equation}
Solving for $\Omega$ then yields the appropriate mimetic factor
\begin{equation}
    \Omega(Y,D,\phi)=\Big (\frac{D}{\sqrt{Y}F(\phi)}\Big )^{2/3}.
\end{equation}
Indeed, this function has conformal weight 2 and the associated mimetic constraint $\Omega(X,P,\phi)=1$ solved for $P$ correctly recovers the original cuscaton Lagrangian:
\begin{equation}
    P=\sqrt{X}F(\phi).
\end{equation}

So far we have considered a k-essence given in an explicit form \eqref{P of X}, however, as we have previously discussed a mimetic theory may support a generalization of k-essence where several k-essence Lagrangians $\mathcal{L}(X,\phi)$ arise in a superposition. This happens since the k-essence Lagrangian arises as a solution of an implicit equation rather than being given explicitly. Fortunately mimetic theory providing this behavior may be engineered as well using the same procedure. We simply start from an implicit equation for the Lagrangian $\mathcal{L}$
\begin{equation}
    \Psi(X,\mathcal{L},\phi)=0,\label{k-essence implicit}
\end{equation}
whose solutions for $\mathcal{L}=\mathcal{L}(X,\phi)$ are locally the k-essence Lagrangians. Our construction can be readily applied. Indeed substituting $P=\mathcal{L}$ and reverting to $Y$ and $D$ through \eqref{mimetic picture variables} results in
\begin{equation}
    \Psi(\Omega^{-1}Y,\Omega^{-2}D,\phi)=0,\label{Omega implicit 2}
\end{equation}
which is qualitatively no different from \eqref{Omega implicit}. Proving that solutions for $\Omega$ locally yield the appropriate mimetic factor is completely analogous with the above argument.

\section{Constraints and radiation breakdown}\label{section: breakdown}
\begin{figure}[ht]
\centering 
\includegraphics[scale=1]{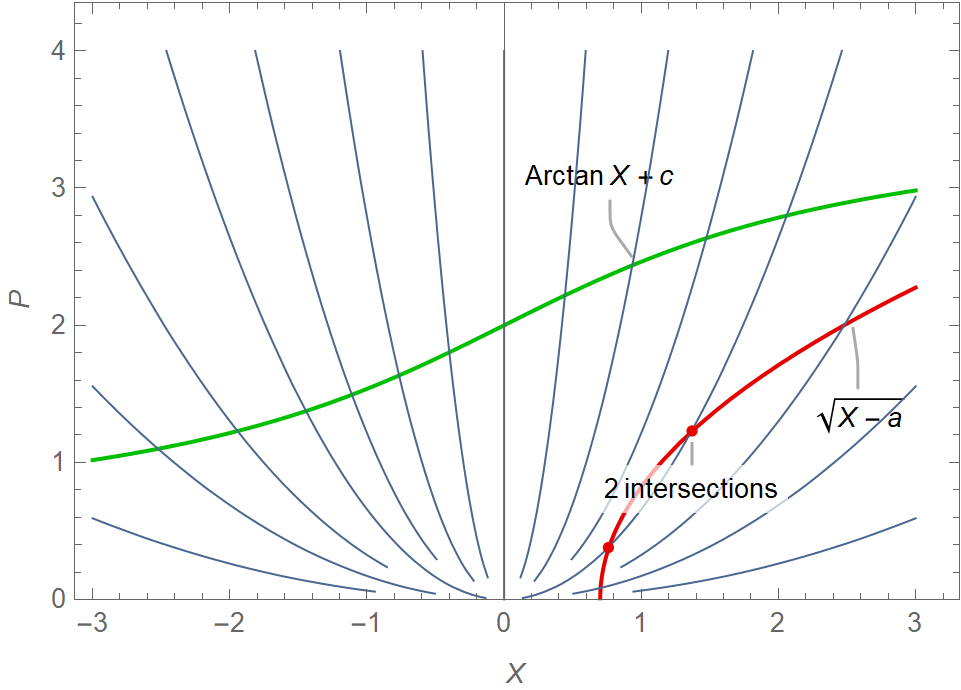}
\caption{\label{fig:1} An example of an admissible k-essence Lagrangian (green), which crosses each arm ($X>0$ or $X<0$) of each parabola at most once. The non-admissible Lagrangian (red) does cross some arms twice. Thus it cannot be fully recovered in a mimetic k-essence theory.}
\end{figure}

The above method might seem arbitrarily general as we have not assumed anything of the k-essence Lagrangian \eqref{P of X}. While there is no problem in finding such implicit equation, it is the need to solve it, which becomes problematic and in some cases impossible. This limits the space of k-essences that are obtainable from mimetic gravity. To demonstrate this, recall how k-essence arises from our theory in the first place. Given a function $\Omega$ we obtain the k-essence Lagrangian $\mathcal{L}(X,\phi)$ (up to an overall scale $Q$) from the associated mimetic constraint $\Omega(X,P,\phi)=1$ as a solution for $P$
\begin{equation}
    P=\mathcal{L}(X,\phi),\label{k-essence curve}
\end{equation}
For a fixed $\phi$ such expression defines a curve in the $X,P$ space. Crucially, the function $\Omega$ is by definition equal to $1$ on this curve. On the other hand, due to the scaling properties \eqref{Omega transformation}, the function $\Omega$ evaluated on any parabola
\begin{equation}
    P=a(\phi)X^{2},\label{gauge orbit}
\end{equation}
has a simple form. For a fixed $a(\phi)$ and $X>0$
\footnote{We consider $X>0$ for the sake of simplicity as only the absolute value of $X$ can be factored out. The general equation valid for all values of $X$ is
\begin{equation*}
    \Omega(X,a(\phi) X^{2},\phi)=|X|\Omega(\mathrm{sgn}(X),a(\phi),\phi).
\end{equation*}
This is due to the gauge parameter being always positive in \eqref{Omega transformation}. As a consequence, \eqref{gauge orbit 2} does not relate the values of $\Omega$ from the left ($X<0$) with the right ($X>0$) arm of the parabola.} we get:
\begin{equation}
    \Omega(X,a(\phi) X^{2},\phi)=X\Omega(1,a(\phi),\phi).\label{gauge orbit 2}
\end{equation}
We see that $\Omega(X,a(\phi)X^{2},\phi)$ is monotonous in $X$. Therefore, $\Omega$ can satisfy the mimetic constraint $\Omega(X,P,\phi)=1$ at most at a single point on each arm of the parabola \eqref{gauge orbit 2}, for any fixed $a$ and $\phi$! In other words, the curve \eqref{k-essence curve} cannot cross any arm of any parabola \eqref{gauge orbit 2} more then once. That is: up to once for $X>0$, up to once for $X<0$ and up to once for $X=0$. This is true for any k-essence that arose from mimetic gravity. This is illustrated in figure \ref{fig:1}. Consequently, if a given k-essence Lagrangian violates this then it cannot be reconstructed through our method.

A second way our reconstruction of k-essences may fail is due to violations of the implicit function theorem in equation \eqref{Omega implicit}\footnote{This is very closely related to the previous criterion. Indeed, if a curve of a k-essence Lagrangian simultaneously fails the above criterion and is continuous and differentiable, then a violation of the implicit function theorem is guaranteed, due to the mean value theorem.}. Interestingly these points have a very clear graphical and physical interpretation. To demonstrate this consider the more general implicit equation \eqref{Omega implicit 2}.
The implicit function theorem for $\Omega$ is violated exactly when
\begin{equation}
\partial_{\Omega}\Psi= 0,\label{Omega implicit function theorem}
\end{equation}
on the constraint surface, which is defined by \eqref{Omega implicit 2}. We are ultimately interested in the properties on the mimetic constraint surface so we can additionally require $\Omega=1$ and relabel $Y,D\rightarrow X,P$. Thus evaluating \eqref{Omega implicit function theorem} under these conditions ($\Psi=0$ and $\Omega=1$) we obtain
\begin{equation}
-\partial_{\Omega}\Psi=\partial_{X}\Psi X+2P \, \partial_{P}\Psi =\mathrm{grad}\ \Psi\cdot (X,2P,0).\label{gauge tangency}
\end{equation}
The right hand side is a dot product of two vectors: the vector $\mathrm{grad}\Psi$, which is a normal to the graph of the resulting k-essence Lagrangian. On the other hand $(X,2P,0)$ is a tangent to the parabolas \eqref{gauge orbit}. Therefore we see that indeed the implicit function theorem is violated exactly when the graph of the k-essence Lagrangian becomes tangent to an arbitrary parabola in the $X,P$ plane.

Interestingly, the violation of the implicit function theorem exactly coincides with the equation of state of the mimetic k-essence becoming that of radiation! Indeed, the $w$ parameter of the mimetic fluid with a Lagrangian $P=\mathcal{L}(X,\phi)$ (in its rest frame) is given as
\begin{equation}
w=\frac{P}{2P_{X}X-P}.\label{EoS}
\end{equation}
Let us focus on a single point $X_{r},P_{r},\phi_{0}$ with $P_{r}=L(X_{r},\phi_{0})$. As long as we keep astray from $X_{r}=0$ we can always find an $a$ such that
\begin{equation}
    P_{r}\equiv a(\phi_{0})X^{2}_{r}.
\end{equation}
Now suppose that the implicit function theorem is violated in the sense of \eqref{Omega implicit function theorem}. As we have shown, this implies tangency to the above parabola. That is
\begin{equation}
    P_{X}=2a(\phi_{0})X_{r}.\label{gauge tangency parabola}
\end{equation}
Plugging this into \eqref{EoS} we obtain
\begin{equation}
w=\frac{a(\phi_{0})X^{2}_{r}}{4a(\phi_{0})X^{2}_{r}-a(\phi_{0})X^{2}_{r}}=\frac{1}{3}.
\end{equation}
The implication goes both ways. Starting with $w=1/3$ and using \eqref{EoS} we could find \eqref{gauge tangency parabola} in the same manner. Thus we find that any violation of the implicit function theorem\footnote{If the violation occurs on the $X$ or $P$ axis the above calculation breaks down and one needs to evaluate \eqref{EoS} through a limiting procedure that may depend on the way the limit is taken.} is characterized by a radiation equation of state.

Any violations of the above considerations causes our method for reconstructing mimetic theory from a k-essence to fail. However, such failure does not manifest in our inability to find a solution of the implicit equation \eqref{Omega implicit}. On the contrary, it leads to existence of multiple solutions. These produce the correct k-essence Lagrangian upon enforcing the mimetic constraint but only for a limited range of values of $X$. Typically the end values of these ranges correspond to the radiation equation of state.
Let us illustrate this with a simple example: consider the free canonical scalar field with a flat potential, characterized by an arbitrary mass scale $m$:
\begin{equation}
    \mathcal{L}=\frac{1}{2}X-\frac{m^{4}}{16}.
\end{equation}
This theory is clearly well defined for all values of $X$. Applying our method we obtain the equation for $\Omega$, a quadratic equation in this case:
\begin{equation}
    m^{4}\Omega^{2}-8Y\Omega+16D=0.
\end{equation}
This yields two solutions:
\begin{equation}
    \Omega_{\pm}=\frac{4}{m^{4}}(Y\pm\sqrt{Y^{2}-m^{4}D}).
\end{equation}
By direct substitution one can check that the mimetic constraint $\Omega_{\pm}(X,P)=1$ is indeed solved by the sought after Lagrangian
\begin{equation}
    P=\frac{1}{2}X-\frac{m^{4}}{16},
\end{equation}
however, for $\Omega_{-}$ this solution is valid only for $X\ge m^{4}/4$. The other solution $\Omega_{+}$ then provides the other branch of the Lagrangian as it is valid only for $X\le m^{4}/4$.

The value $X=\frac{m^{4}}{4}$, where the solutions branch corresponds to the equation of state of radiation
\begin{equation}
    w|_{X=\frac{m^{4}}{4}}=\frac{1}{3}.
\end{equation}



\section{Revisiting Weyl symmetry}\label{section: non-Weyl}
Till now we have assumed that ansatz \eqref{mimetic ansatz} is invariant under the Weyl transformations of the auxiliary metric $h_{\mu\nu}$. This was ensured by requiring that the function $\Omega$ satisfies the relations \eqref{Omega transformation}. In our recent paper \cite{Jirousek:2022rym} we have shown that in the simpler case of mimetic gravity, featuring only $Y$ and $\phi$, such assumption is not needed in order to recover mimetic gravity. Indeed, integrating out the auxiliary conformal degree of freedom effectively transforms a general function $\Omega(Y,\phi)$ into a linear function of $Y$ as is needed for the mimetic ansatz \eqref{mimetic DM ansatz}. The same mechanism is applicable in the present setup as well. To demonstrate this we consider an ansatz
\begin{equation}
    g_{\mu\nu}=h_{\mu\nu}\,\Omega(Y,D,\phi).
\end{equation}
In contrast to \eqref{mimetic ansatz} we put no extra assumptions on the function $\Omega$. The above expression has the form of a conformal transformation of the auxiliary metric $h_{\mu\nu}$ regardless of the details of $\Omega$. Therefore, we can repeat the steps \eqref{action expanded}-\eqref{field redefinition phi} to arrive at the action:
\begin{equation}
    S[g,\theta,\phi,A_{\mu},\lambda]=-\frac{1}{2}\int d^{4}x\sqrt{-g}\bigg [R(g) +\tilde{\lambda}\big (\theta^{-1}\Omega(\theta X,\theta^{2}P)-1\big )\Bigg].\label{chi action}
\end{equation}
A crucial point here is that the field $\theta$ does not drop out of this action. This has a very simple interpretation: $\theta$ carries the information about the conformal degree of freedom of $h_{\mu\nu}$. The Weyl invariance of the original theory was due to the absence of this mode in the action. In this sense, the Weyl symmetry of the original setup was "fake"  \cite{TSAMIS1986457,Jackiw:2014koa,Oda:2016pok}. Conversely, breaking of this invariance here is due to its presence. Consequently, on top of the constraint
\begin{equation}
    \theta^{-1}\Omega(\theta X,\theta^{2}P)=1,\label{no weyl constraint}
\end{equation}
we obtain a non-trivial equation of motion associated with $\theta$:
\begin{equation}
    \Psi(\theta X,\theta^{2}P,\phi)\equiv\Omega(\theta X,\theta^{2}P)-\Omega_{X}(\theta X,\theta^{2}P)\theta X-2\Omega_{P}(\theta X,\theta^{2}P)\theta^{2}P=0.\label{implicit chi}
\end{equation}
Since $\theta$ entered with no derivatives in \eqref{implicit chi}, this equation of motion is purely algebraic in $\theta$. This, in principle, allows us to solve for $\theta$ in terms of $X,P$ and $\phi$ as $\theta(X,P,\phi)$. However, equation \eqref{implicit chi} may admit solutions that do not determine $\theta$ at all. Instead they introduce an additional constraint on $X,P$ and $\phi$. These two options lead to a very different behaviors of the resulting theory.\

Let us first focus on the former case - when $\theta$ can be solved for algebraically. We notice that $\theta$ enters \eqref{implicit chi} in a very particular manner: only in combinations $\theta X$ and $\theta^{2}P$. This is essentially the same way $\Omega$ entered equation \eqref{Omega implicit 2} (up to an inverse power). This particular form resulted in solutions for $\Omega(Y,D,\phi)$ having a conformal weight $2$ under Weyl transformations of $h_{\mu\nu}$. Repeating the same argument reveals that any solution $\theta(X,P,\phi)$ will have a conformal weight $-2$ under Weyl transformations of $g_{\mu\nu}$:
\begin{equation}
    \theta(\omega^{-2}X,\omega^{-4}P,\phi)=\omega^{2}\theta (X,P,\phi).
\end{equation}
Since this solution has been determined algebraically from an equation of motion associated to $\theta$, we are allowed to plug it back into the Lagrangian \eqref{chi action} \cite{Pons:2009ch}. This way $\theta$ is integrated out of the action. The constraint term in \eqref{chi action} becomes
\begin{equation}
    -\frac{1}{2}\int d^{4}x\sqrt{-g}\tilde{\lambda}(\tilde{\Omega}(X,P,\phi)-1),\label{mimetic constraint 2}
\end{equation}
where we have introduced
\begin{equation}
    \tilde{\Omega}(X,P,\phi)\equiv\theta^{-1}(X,P,\phi)\Omega\big(\theta(X,P,\phi) X,\theta^{2}(X,P,\phi)P\big).
\end{equation}
Notice that, due to the particular conformal weight of $\theta$, the entire function $\tilde{\Omega}$ has a conformal weight $2$. Therefore, the resulting theory has the same form as \eqref{action mimetic constraint} and is classically indistinguishable from a mimetic theory obtained from an ansatz
\begin{equation}
    g_{\mu\nu}=h_{\mu\nu}\,\tilde{\Omega}(Y,D,\phi).
\end{equation}
Hence we see that even when we start with a general conformal transformation of $h_{\mu\nu}$ the theory generates Weyl symmetry dynamically. Consequently, our discussion from the previous sections of this paper is applicable to these cases as well.\

Let us move on to the second type of solutions - when \eqref{implicit chi} does not determine $\theta$. Instead it introduces a novel constraint on the fields $X,P$ and $\phi$. Due to the way $\theta$ enters its equation of motion, this constraint has a very specific form.
\begin{equation}
    P=X^{2}c(\phi).\label{no weyl parabola}
\end{equation}
We can show this by solving \eqref{implicit chi} for its second argument $\theta^{2}P$. Writing this solution formally yields a general form
\begin{equation}
    \theta^{2}P=c(\theta X,\phi).
\end{equation}
where $c(\theta X,\phi)$ is an undetermined function. By our assumption this equation must be independent of $\theta$. This occurs only when $c$ depends quadratically on its first argument and we can divide $\theta$ out\footnote{Recall that throughout this paper we assume $\theta\neq 0$.} - obtaining \eqref{no weyl parabola}. Note that this expression admits solutions of the type $P=0$. Repeating the same reasoning for the first argument $\theta X$ (instead of $\theta^{2}P$) reveals that $X=0$ is also an admissible solution, which may be viewed as a limit $c(\phi)\rightarrow\infty$ in equation \eqref{no weyl parabola}.\

Our assumptions purposely made us fail to determine $\theta$ from its own equation of motion. One might hope to obtain it from the constraint \eqref{no weyl constraint} instead, however, this is generally impossible. The relation between the two constraints \eqref{implicit chi} and \eqref{no weyl constraint} is
\begin{equation}
    \Psi(\theta X,\theta^{2}P,\phi)=-\theta^{2}\partial_{\theta}\Big (\theta^{-1}\Omega(\theta X,\theta^{2}P)\Big ).
\end{equation}
This implies that the implicit function theorem for $\theta$ in \eqref{no weyl constraint} is violated when the $\theta$ equation of motion is satisfied. Therefore $\theta$ cannot be determined from this equation either and \eqref{no weyl constraint} represents a second constraint on $X,P$ and $\phi$.\footnote{These constraints may happen to be incompatible - signaling inconsistency of the branch.} These may be used to solve for $X$ and $P$ in terms of $\phi$.
\begin{align}
    X&=b(\phi),\\
    P&=\frac{c(\phi)}{b^{2}(\phi)},\label{no weyl solutions}
\end{align}
where $c(\phi)$ comes from \eqref{no weyl parabola} and $b(\phi)$ is an undetermined function. These dynamics are considerably different from the k-essence behavior that we have seen throughout this paper. Inspecting the above equations we can immediately see that $\phi$ is determined through a first order equation, which is essentially an example of mimetic dark matter constraint. This is fundamentally different from a generic k-essence where $\phi$ is (up to a few exceptions) a proper propagating degree of freedom satisfying a second order equation of motion. To identify what kind of dynamics this theory describes we look at the equations of motion. The gravitational equations
\begin{equation}
    G_{\mu\nu}+\tilde{\lambda}\Omega_{X}(\theta X,\theta^{2}P)\partial_{\mu}\phi\partial_{\nu}\phi+\frac{\tilde{\lambda}\theta}{2}\Omega_{P}(\theta X,\theta^{2}P)Pg_{\mu\nu}=T_{\mu\nu},\label{no weyl einstein}
\end{equation}
are still those of GR with an additional perfect fluid component with velocity potential $\phi$. It's energy density and pressure are
\begin{align}
    \rho &= -\tilde{\lambda}\Omega_{X}(\theta X,\theta^{2}P)X-\frac{\tilde{\lambda}\theta}{2}\Omega_{P}(\theta X,\theta^{2}P)P,\\
    p &= \frac{\tilde{\lambda}\theta}{2}\Omega_{P}(\theta X,\theta^{2}P)P.
\end{align}
The pressure term can be significantly simplified using the equation of motion for $A_{\mu}$, which yields
\begin{equation}
    \tilde{\lambda}\theta\Omega_{P}(\theta X,\theta^{2}P)=Q=const.
\end{equation}
where $Q$ is a constant of integration. Furthermore plugging in the solutions \eqref{no weyl solutions} allows us to write:
\begin{align}
    \rho &= -\tilde{\lambda}\Omega_{X}(\theta X,\theta^{2}P)X-\frac{Q}{2}P(\phi),\\
    p &= \frac{Q}{2}P(\phi).
\end{align}
We see that the pressure term depends on $\phi$ alone on the equations of motion. It is then natural to interpret it as a potential term for $\phi$. The only difference being the presence of the overall scale $Q$ that enters as a constant of integration - a global degree of freedom. In order to keep track of it we will keep it separate and introduce a potential
\begin{equation}
    V(\phi)=-\frac{1}{2}P(\phi)=-\frac{1}{2}\frac{c(\phi)}{b^{2}(\phi)}.
\end{equation}
The equation \eqref{no weyl einstein} written in terms of $\rho$ and $V$ is
\begin{equation}
    G_{\mu\nu}=(\rho-QV)u_{\mu}u_{\nu}+QV(\phi)g_{\mu\nu}+T_{\mu\nu},
\end{equation}
which corresponds exactly to mimetic dark matter with potential $V(\phi)$ \cite{Lim:2010yk,Chamseddine:2014vna} with an additional global degree of freedom $Q$. The dynamics of $\rho$ are then determined by the equation of motion for $\phi$ which takes the expected form
\begin{equation}
    \nabla^{\mu}\Big [(\rho-QV(\phi))\partial_{\mu}\phi\Big]=-QV'(\phi)\sqrt{b(\phi)}.
\end{equation}
Interestingly, in this branch we can recover both mimetic dark matter \cite{Chamseddine:2013kea} and Henneuax-Teitelboim unimodular gravity \cite{Henneaux:1989zc,Jirousek:2018ago,Hammer:2015pcx} by a single conformal transformation of the metric. This can be achieved via a very simple ansatz\footnote{This is by far not unique. In fact any consistent ansatz that falls into this branch of solutions and is $\phi$- independent will lead to this behavior.}
\begin{equation}
    g_{\mu\nu}=h_{\mu\nu}\,(Y+D-Y^{2})\,.\label{unification}
\end{equation}
The equation of motion for $\theta$ \eqref{implicit chi} and the constraint equation \eqref{no weyl constraint} read respectively
\begin{align}
    \theta^{2}(P-X^{2})&=0,\\
    X+\theta(P-X^{2})&=1.
\end{align}
Solving these gives us exactly the constraint equations of mimetic dark matter and Henneaux-Teitelboim unimodular gravity
\begin{align}
    X=1,\\
    P=1.
\end{align}
The gravitational equation and the equation of motion for $\phi$ yield the expected result
\begin{align}
    G_{\mu\nu}+\tilde{\lambda}\partial_{\mu}\phi \partial_{\nu}\phi+Qg_{\mu\nu}&=T_{\mu\nu},\\
    \nabla^{\mu}\Big (\tilde{\lambda}\partial_{\mu}\phi\Big )&=0,
\end{align}
while $Q$ is a constant of integration obtained from the equation of motion for $A_{\mu}$. Thus we indeed have a system with mimetic dark matter and a cosmological constant given as a constant of integration!


\section{Conclusions and discussion}
In this paper we considered a modification of the mimetic gravity scenario, in which the conformal degree of freedom is isolated using a combination of terms - the kinetic term for a scalar field \eqref{kinetic term} and the Chern-Pontryagin invariant of a Yang-Mills gauge field \eqref{Pontryagin invariant}. On their own, these terms produce simple models of the components of the dark sector - dark matter \cite{Chamseddine:2013kea} and dark energy \cite{Jirousek:2018ago,Hammer:2020dqp} respectively. By combining them in this framework one can in a way interpolate between the two. Interestingly, we found that the result of such interpolation is on-shell equivalent with standard general relativity accompanied by a k-essence scalar field. The associated k-essence Lagrangian arises in this setting as a solution of an implicit equation \eqref{P of X}, and as a result single mimetic theory may support multiple different Lagrangians. This opens the possibility of superpositions of k-essences with varying dynamics, tunneling effects or possibly phase transitions. Furthermore, the theory contains an extra global degree of freedom, or a global charge, that acts as an overall scale of the mimetic k-essence \eqref{mimetic k essence}. The mechanism in which this charge appears is very similar to \cite{Jirousek:2018ago,Jirousek:2020vhy}. Therefore, it is reasonable to expect that this will also result in non-trivial commutation relations with  global quantity \eqref{conjugated} in the theory and consequently in the quantum fluctuations of this energy scale. 
The value of this constant most probably has the origin in quantum cosmology. 

Interestingly, we found that our mimetic k-essences can reproduce almost any k-essence or superfluid in a Weyl-invariant setup. Though, we demonstrate that the theory breaks down when the equation of state of the mimetic fluid becomes ultra-relativistic. This prevents any classical transitions between $w>1/3$ and $w<1/3$. We provided a general method for constructing a mimetic theory that reproduces a given k-essence or even their superpositions and determined the conditions under which this construction is possible.

Finally, we investigated the role of the underlying Weyl invariance in our setup. By abandoning this requirement, the conformal mode of the auxiliary metric $h_{\mu\nu}$ reappears in the action and yields an additional equation of motion. Usually this equation allows us to integrate the new mode out of the action, restoring Weyl symmetry in the process. In light of this, the original Weyl symmetry of the theory seems to be unnecessary. However, in contrast to \cite{Jirousek:2022rym}, there are certain cases in which the mode cannot be integrated out and instead acts as a Lagrange multiplier. In that case, the theory reduces to mimetic dark matter with a potential as was suggested in \cite{Lim:2010yk,Chamseddine:2014vna}. The extra global degree of freedom remains and provides the scale for the given potential. Thanks to this we were able to recover both mimetic dark matter \cite{Chamseddine:2013kea} and Henneaux-Teitelboim unimodular gravity \cite{Henneaux:1989zc} through a single conformal transformation \eqref{unification} of the metric. 

It would be very interesting to develop full scale Dirac Hamiltonian analysis for such mimetic mix. Another interesting avenue to explore is related to boundary conditions and boundary terms in this setup. We leave these problems for a future work. 
\section{Acknowledgements}

Our collaboration is supported by the Bilateral
Czech-Japanese Mobility Plus Project JSPS-21-12 (JPJSBP120212502). This project originated when P. J. and A. V. were enjoying very warm hospitality of the
cosmology group at the Tokyo Institute of Technology.
This productive visit was possible thanks to the JSPS
Invitational Fellowships for Research in Japan (Fellowship ID:S19062) received by A. V. M. Y. would also like to thank the CEICO, Institute of Physics of the Czech Academy of Sciences, for their hospitality during the final stages of this work.
The work of P.~J. was supported by the Grant Agency of the Czech Republic (GAČR grant 20-28525S), through the most of the project. P.~J. also acknowledges funding from the South African Research Chairs Initiative of the Department of Science and Technology and the National Research Foundation of South Africa in the final stages of preparation of the manuscript. K.~S. was supported by JSPS KAKENHI Grant Number JP20J12585 for the initial stages of this work. A.~V. acknowledges
support from the European Structural and Investment Funds and the Czech Ministry of Education, Youth and Sports (Project CoGraDS -CZ.02.1.01/0.0/0.0/15003/0000437).  M.~Y. acknowledges financial support from JSPS Grant-in-Aid for Scientific Research No. JP18K18764, JP21H01080, JP21H00069.

\bibliographystyle{utphys}
\bibliography{references}

\end{document}